\documentclass[journal]{IEEEtran}
\usepackage{stfloats}
\usepackage{blindtext}
\usepackage{multicol}
\usepackage[utf8]{inputenc}
\usepackage{csquotes}
\usepackage{enumitem}
\usepackage{amsmath}
\usepackage{amssymb}
\usepackage{graphicx}
\usepackage{diagbox}
\usepackage{subcaption}
\usepackage{makecell}
\usepackage{bm}
\usepackage{algorithm}
\usepackage[noend]{algpseudocode}
\def\tablename{Table}

\title{From Multi-agent to Multi-robot: A Scalable Training and Evaluation Platform for Multi-robot Reinforcement Learning}

\author{
Zhiuxan Liang, Jiannong Cao, \textit{Fellow, IEEE}, Shan Jiang, Divya Saxena, Jinlin Chen, Huafeng Xu
\thanks{Zhixuan Liang, Jiannong CaO, Shan Jiang, Divya Saxena, Jinlin Chen, and Huafeng Xu were with Department of Computing, The Hong Kong Polytechnic University, Hong Kong SAR.
Email: zhixuan.liang@connect.polyu.hk}
}

\begin{document}

\maketitle

\begin{abstract}
Multi-agent reinforcement learning (MARL) has been gaining extensive attention from academia and industries in the past few decades. One of the fundamental problems in MARL is how to evaluate different approaches comprehensively. Most existing MARL methods are evaluated in either video games or simplistic simulated scenarios. It remains unknown how these methods perform in real-world scenarios, especially multi-robot systems. This paper introduces a scalable emulation platform for multi-robot reinforcement learning (MRRL) called SMART to meet this need. Precisely, SMART consists of two components: 1) a simulation environment that provides a variety of complex interaction scenarios for training and 2) a real-world multi-robot system for realistic performance evaluation. Besides, SMART offers agent-environment APIs that are plug-and-play for algorithm implementation. To illustrate the practicality of our platform, we conduct a case study on the cooperative driving lane change scenario. Building off the case study, we summarize several unique challenges of MRRL, which are rarely considered previously. Finally, we open-source the simulation environments, associated benchmark tasks, and state-of-the-art baselines to encourage and empower MRRL research.
\end{abstract}


\section{Introduction}
Deep reinforcement learning (DRL) has emerged as a promising approach for decision-making problems, widely adopted in various domains such as finance \cite{Avraam2021Price}, healthcare \cite{Matthew2021Improving}, recommendation \cite{Yu2019Interactive}, and intelligent transportation systems \cite{Neetesh2021Fuzzy}. Recent advances in DRL have shifted the research focus from a single agent to multiple agents in accomplishing complex and complicated tasks \cite{fan2020distributed}\cite{lin2018efficient}. Hence, multi-agent reinforcement learning (MARL) arises as a new research direction \cite{vinyals2017starcraft}. In MARL, multiple agents sense and act to perform tasks in shared environments \cite{zheng2019wuji}\cite{zhou2019multi}. At each timestep, the agents observe the environment, select optimal actions, and receive rewards (feedback signal) related to task accomplishment. MARL has been receiving remarkable attention in the research communities and industries because of its great potentials in real-world applications.

\begin{figure}[!ht]
\centering
\includegraphics[width=\linewidth]{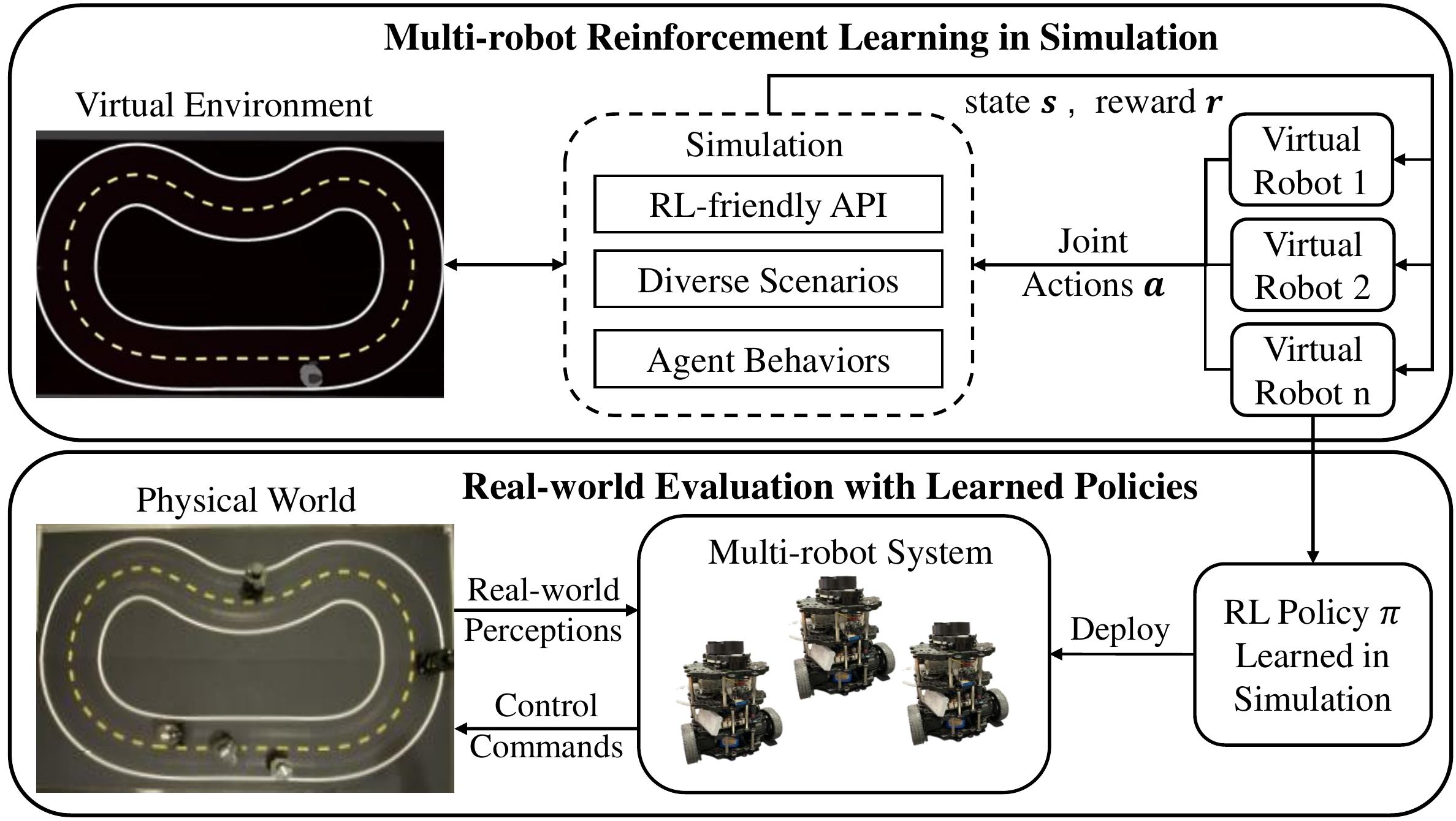}
\caption{Training with different reinforcement learning algorithms in the simulation and evaluating them in the real-world multi-robot system.}
\label{fig:1}
\end{figure}

One of the fundamental problems of MARL lies in real-world performance evaluation because most MARL models are trained in simulated environments. For example, Lowe et al. proposed a multi-agent actor-critic method and evaluated it in a 2D space environment where the agents are abstracted as circle entities with navigation capability to different landmarks without collisions \cite{lowe2017multi}\cite{foerster2017counterfactual}. The state space of each agent is simplified as an absolute position of the agent itself and the relative positions of other agents. The action space is defined as the moving directions and speeds. Besides, several 3D gaming benchmark environments are developed for training and evaluation, such as StarCraft Multi-Agent Challenge (SMAC) \cite{vinyals2017starcraft} and hide-and-seek \cite{baker2019emergent}. Nevertheless, these platforms are limited in the complexity of interactions and physics simulation. At the same time, real-world robots provide an appealing domain for evaluating reinforcement learning algorithms. Robots learn adaptive control strategies from low-level sensor observations, referred to as multi-robot reinforcement learning (MRRL).

However, building an MRRL platform is more than challenging. First, direct training in real-world environments is expensive because exploring strategy learning can lead to undesired damages. Second, the high dimensional state space (e.g., image and point cloud data) and action space (e.g., continuous speeds) require more training to learn the optimal control strategy. Beyond the cost of training, software and hardware components design is a touch engineering problem because there are many choices in algorithm implementation and robot setup. Still, there are some real-world multi-robot platforms such as multi-robot pathfinding platforms \cite{pickem2017robotarium} and multi-robot pattern formation platforms \cite{wang2019pattern}. However, these platforms are not MRRL-friendly to the training and evaluation. Furthermore, the application of MARL in the multi-robot system and the gap between simulation and reality are still unclear and demand investigations.

\begin{figure*}[!ht]
	\centering
	\includegraphics[width=.95\linewidth]{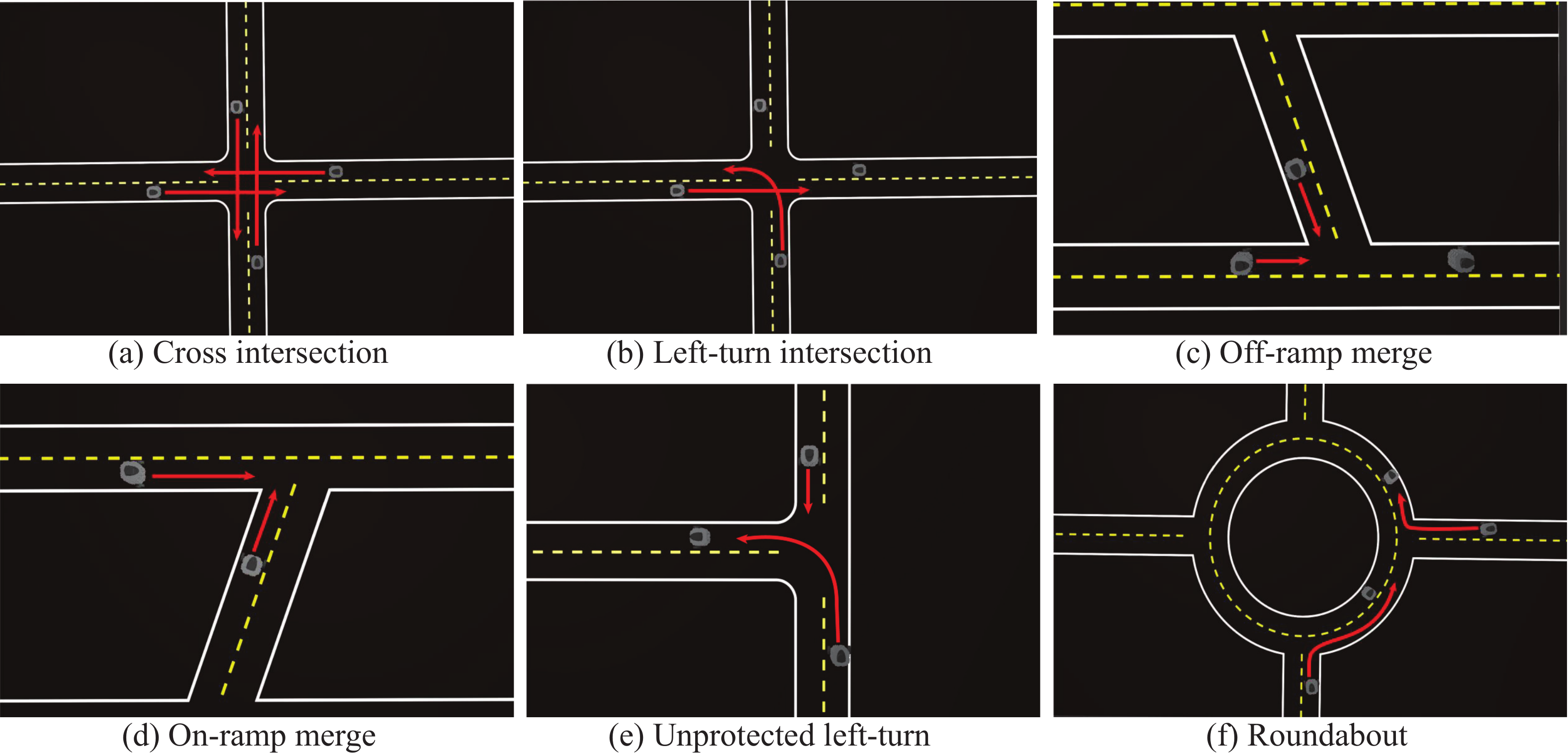}
	\caption{Selective scenarios of multi-robot interaction in SMART. Different scenarios involve different environments and heterogeneous robots.}
	\label{fig:2} 
\end{figure*}

This paper introduces a scalable multi-robot reinforcement learning platform (SMART) for training and evaluation to address the challenges above. First, we develop a realistic simulation environment with diverse interaction scenarios. Specifically, we consider the multi-robot coordination in the autonomous driving scenario, which is one of the most promising applications of MRRL. Instead of directly training in the real world, SMART enables pre-training in the simulations, avoiding unfavorable robotic damages and time consumption. In this way, we can simulate diverse intersection scenarios with various learned agents and social agents controlled by the proposed behavior models. Second, we introduce agent-environment APIs designed for easy-to-use implementation and evaluation. Besides, we have developed a real-world testbed corresponding to simulated scenarios. Finally, we conduct a case study on two lane-merge scenarios, extend several state-of-the-art MARL models to multi-robot systems, and discuss the gap between simulation and reality. The main contributions of this paper are summarized as follow:
\begin{itemize}
    \item We discuss the training and evaluation environments in MRRL and introduce a novel platform that includes the simulation environment and real-world platform as shown in Fig.~\ref{fig:1}.
    \item Our platform provides a simulative environment with different multi-robot interaction scenarios and MRRL-friendly APIs for training.
    \item To investigate the gap between simulation and reality, we develop a multi-robot system as the real-world platform and conduct a case study on two lane-merge scenarios.
    \item Building off the case study, we discuss the challenges of applying deep reinforcement learning in multi-robot systems, which is rarely considered in reinforcement learning research that focuses only on simulated domains.
\end{itemize}

The rest of this paper is organized as follows. In Sec.~\ref{sec:platform}, we discuss the design goal of our platform and the corresponding implementation. Sec.~\ref{sec:case-study} gives an example study on multi-robot cooperative lane merge scenarios and analysis of the experiment results and discusses several lessons we learn in Sec.~\ref{sec:lesson}. In Sec.~\ref{sec:related-work}, we present some related works. Finally, Sec.~\ref{sec:conclusion} concludes this paper. 

\section{SMART: Scalable Multi-robot\\Reinforcement Learning Platform}\label{sec:platform}

In this section, we introduce a SMART, i.e., a \underline{s}calable \underline{m}ulti-robot le\underline{ar}ning pla\underline{t}form, which provides training services in the simulation environment and evaluation in the real-world multi-robot system. We first present the design goals of SMART and then introduce the system architecture and implementation details. Finally, we illustrate the support from SMART through different stages when designing new MRRL algorithms.

\subsection{Design Goals of SMART} 

\textbf{Seamless Integration of Simulation Environments and Real-world Platform.} 
MRRL requires the robots to learn from intensive interactions in shared environments during the training phase. It is inefficient to let the robots learn directly in the real world. Training a reinforcement learning model demands numerous data while the trial-and-error process takes considerable time and human efforts to collect even a tiny portion. Furthermore, trial and error can be costly and even risky because the robots can be damaged, leading to unpredictable dangers to the environments and even experimental staff.

In SMART, we aim to provide a realistic simulation environment for the pre-training services. SMART is realistic to mimic the real world, facilitating the evaluation of various MRRL approaches. Meanwhile, it is also necessary to build up a real-world platform considering the gap between simulation and reality. In this way, different MRRL approaches can be pre-trained in the simulator and then evaluated in the platform.

\textbf{Support of diverse Interaction Scenarios.} One fundamental characteristic in MRRL is the \textit{mutual influence} among robots such as multi-robot cooperative driving in different lane merge scenarios. The actions of all robots determine the state transition and reward function. To investigate the mutual influence among robots, we design diverse interaction scenarios involving the coordination of multiple robots. In these scenarios, the state spaces and the cooperation strategy among robots can differ significantly. Hence, the diverse interaction scenarios can comprehensively evaluate different MRRL methods.

\begin{figure*}[!ht]
	\centering
	\includegraphics[width=0.65\linewidth]{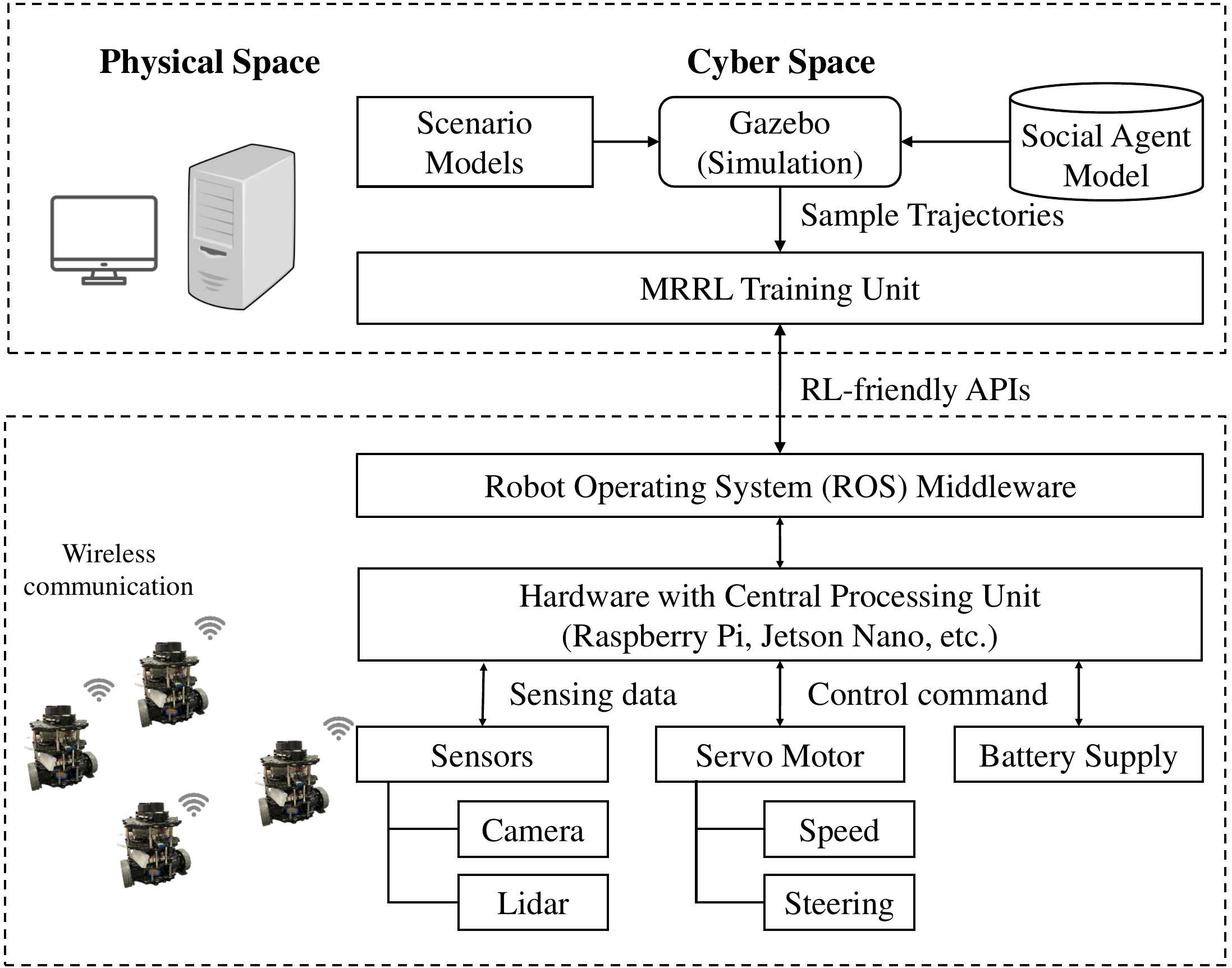}
	\caption{System architecture of SMART, including the design of cyber and physical spaces}
	\label{fig:3} 
\end{figure*}

\textbf{Providing MRRL-friendly APIs.} The MRRL mechanism differs a lot compared to traditional machine learning approaches relying on data provided by users.
At each step, the robots observe the environment, perform actions, receive the team reward, and then repeat.
Such a unique interaction mechanism is also called the robot-environment cycle. Considering the algorithm implementation, we provide corresponding interacting APIs that are easy to use for the MRRL researchers.

\textbf{Support for Heterogeneous Robots.} Real-world applications often involve the cooperation of heterogeneous robots. In SMART, we provide two types of robots. One type of robot is controlled by the MRRL models, called an intelligent robot. Another type of robot, called a social robot, is controlled by the rule-based model or random policy. We argue that the interaction among heterogeneous robots will lead to new challenges for different MRRL approaches. To increase the diversity of robots, the state space and action space are different and configurable by the users. Researchers can specify them according to the assumption of their algorithms.

\textbf{Providing Baseline Libraries.} 
Major enterprises or organizations provide some popular and open-source reinforcement learning baselines, such as OpenAI Gym \cite{brockman2016openai} or Deepmind Acme \cite{hoffman2020acme}. However, few of them provide the baselines of MRRL. To facilitate the reproduction and comparison of other algorithms, we also include different standard MRRL algorithms in SMART. These algorithms are implemented with popular deep learning frameworks, such as Tensorflow and Pytorch, so that the researchers can quickly repeat, refine, and identify new ideas.

\subsection{System Architecture}
To achieve the design goals, we propose the overall system architecture of SMART as shown in Fig.~\ref{fig:3}. It mainly consists of a simulation environment and a real-world platform. The simulation is built on the top of the Robot Operating System (ROS) \cite{quigley2009ros}. The physics simulation is provided by Gazebo \cite{koenig2004design}, a famous physics engine in ROS. SMART provides the world model and the control of robots. Besides, we also provide friendly learning APIs for the training. After pre-training in the simulation, we develop a multi-robot system as the real-world platform to evaluate the learned policies. Each robot in the platform can sense the environments and other robots, communicate with others, and make individual decisions.

\subsection{Software Implementation}

\textbf{Software.} The main component of the software in SMART is the highly realistic simulation environments, providing the primary training and evaluation services to prototype, test, and rectify the algorithms. In this paper, we study cooperative driving as the major scenario, which has been widely studied in multi-robot cooperation \cite{bhalla2020deep}\cite{jingliang2019hierarchical}\cite{chen2019attention}. To meet the requirements of complex interactions, we design different environment models, including various interaction scenarios. Fig.~\ref{fig:2} shows the selective scenarios in our platform. These environment models can be configured and loaded in the simulator. We argue that a powerful MRRL approach can perform well in complex interaction scenarios. Also, we provide an option with different types of social robots in the environment. These robots are controlled by simple scripts or rules provided by the users. Before the training, users can select and configure the environment and social robots through the provided interface.

\begin{figure}[!ht]
	\centering
	\includegraphics[width=\linewidth]{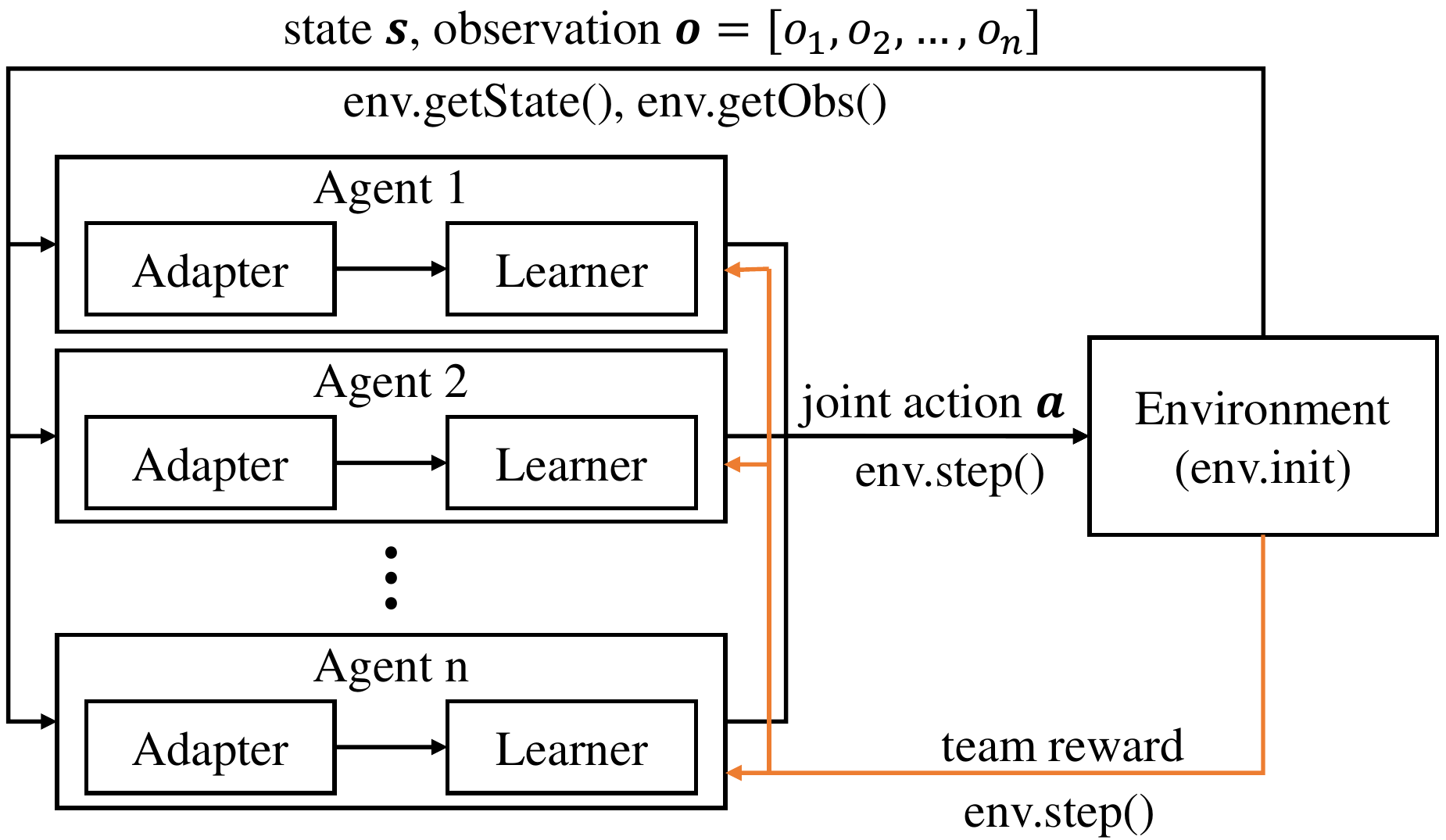}
	\caption{The API design of SMART is inspired from the agent-environment interaction loop, which is programming-friendly to reinforcement learning researchers.}
	\label{fig:api-design} 
\end{figure}

\textbf{Robot-environment APIs.} Different from the training of deep learning models, the training procedures of reinforcement learning models require frequent interaction between the robots and environments. Therefore, we design specific robot-environment interacting APIs for easy-to-use implementation and training. For example, the training can start with the \textit{reset} interface, which resets the environment and returns the observations of all robots. Then each robot can choose an action and respond to the environment through the \textit{step} interface and obtain the rewards. These interfaces are natural for understanding and convenient for implementing different algorithms. The interfaces for MRRL are summarized as follows:

\begin{itemize}
    \item Env.\textit{\textbf{make}(senario=scenario\_name)}: create the learning environment with specified scenario name.
    \item Env.\textit{\textbf{reset()}}: reset the environment and return the observation of all robots.
    \item Env.\textit{\textbf{step}(actions)}: execute the actions generated by different MRRL algorithms and return the next observations, rewards of all robots.
\end{itemize}

Considering the deployment on the real-world multi-robot system, implementing these interfaces is incorporated with some essential services provided by ROS. Therefore, training in the simulation and evaluation in the real-world platform can share the same elegant APIs. More details about the implementation are discussed in Sec.~\ref{sec:case-study}.

\begin{figure}[!ht]
\centering
\includegraphics[width=\linewidth]{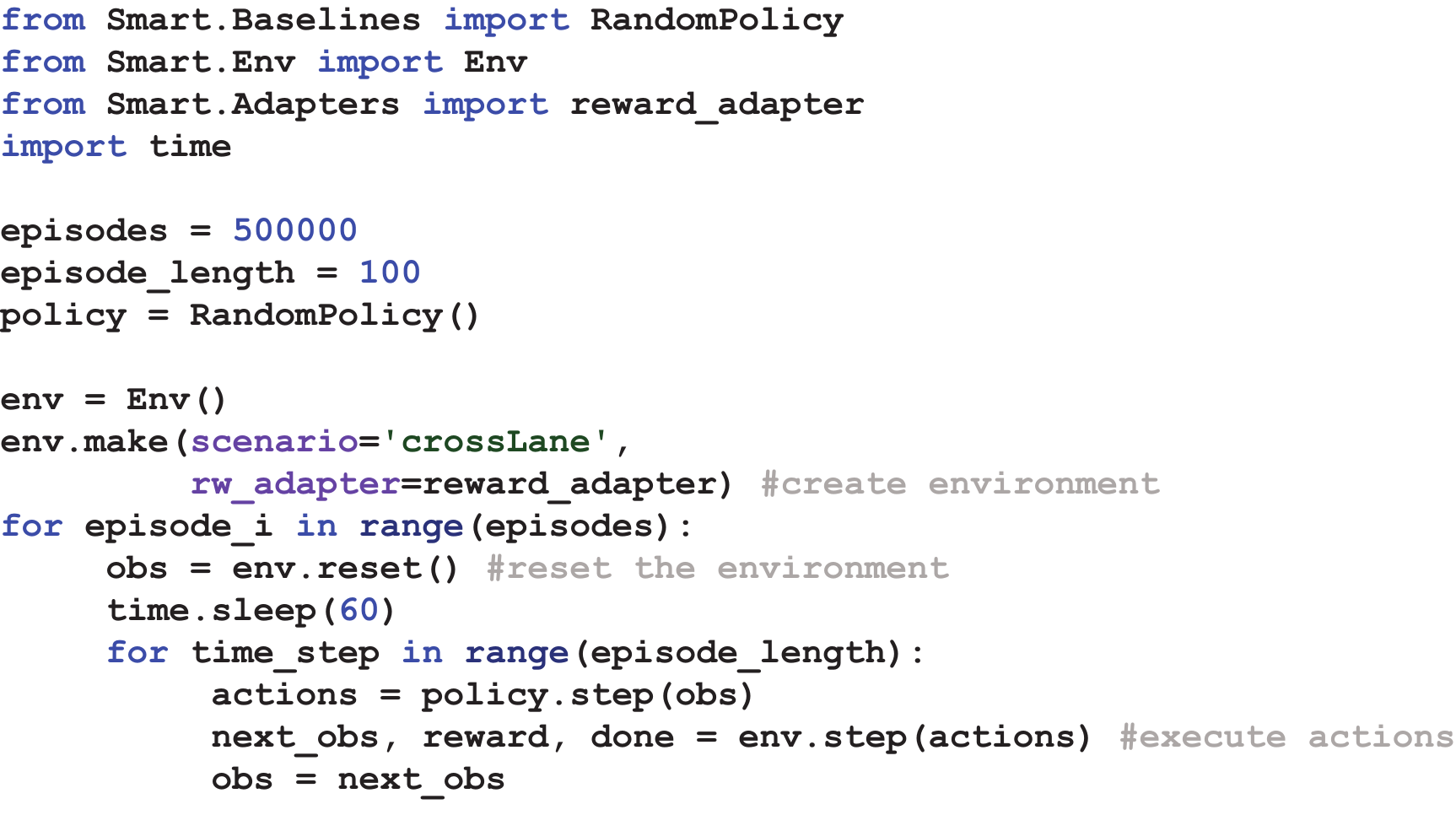}
\caption{An example of creating environment and training}\label{fig:code}
\end{figure}

\subsection{Social Agent Model}

\begin{figure}[!ht]
\centering \includegraphics[width=0.5\linewidth]{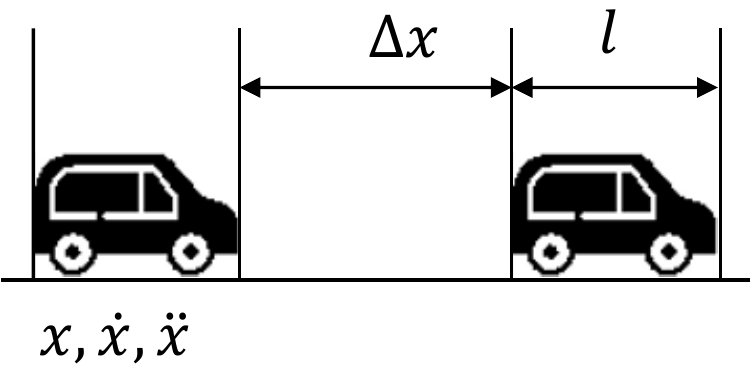}
\caption{Intelligent driver model}
\label{fig:IDM-model}
\end{figure}

A real-world environment usually involves intelligent agents and other social agents controlled by humans. However, the driving behavior of humans is complex and related to their perception and driving preference. Intelligent driving model (IDM) is considered to be the simplest complete accident-free model, which can generate real vehicle acceleration curve and reasonably describe vehicle behavior 
traffic simulation. The parameters of the intelligent driving model are very intuitive, so that it can describe the driver's behavior and establish a limitation model for the physical performance of the vehicle. In this model, the acceleration speed is defined as:

\begin{center}
    $\ddot{x}=a\left(1-\left(\frac{v}{v_{f}}\right)^{4}-\left(\frac{\Delta S^{*}(v, \Delta \mathrm{v})}{\Delta x}\right)^{2}\right)$
\end{center}

where $\frac{v}{v_{f}}$ compares the difference between the current speed and expected speed, while $\Delta S^{*}(v, \Delta \mathrm{v})$ is the expected distance calculated as follow:
\begin{center}
    $\Delta S^{*}(v, \Delta \mathrm{v})=S_{0}+\max \left(v \cdot \Delta t+\frac{v \cdot \Delta \mathrm{v}}{2 \sqrt{a b}}, 0\right)$
\end{center}

where $s_0$ is the safe distance, $\Delta t$ is the reaction time, $a$ is the vehicle's maximum acceleration speed, and $b$ is the vehicle's moderate deceleration. 

\begin{table*}[!ht]
\centering
\caption{Hardware comparison of Smartbot-alpha and Smart-beta}
\begin{tabular}{|c|c|c|}

\hline 
 & Smartbot-alpha & Smartbot-beta\\
\hline 
Mainboard & Raspberry pi 3b+ & Nvidia Jetson Nano\\
\hline
Motion Control Borad & OpenCR1.0 & OpenCR1.0\\
\hline
Camera & RPI Camera G ($500$ pixels and $160$ degree) & Sony IMX219 ($800$ pixels and $170$ degree) \\
\hline 
LDS(Laser Distance Sensor) & $360$-degree LDS-01 & $360$-degree LDS-01\\
\hline
Actuator & XL430-W250 & XM430-W210\\
\hline
Battery & $11.1$V $1800$mAh / $19.98$Wh $5$C & $11.1$V $3600$mAh / $19.98$Wh $5$C\\
\hline
Maximum translational velocity & $0.22$m/s & $0.26$m/s\\
\hline
Maximum rotational velocity & $2.84$rad/s & $1.82$rad/s\\
\hline
\end{tabular}
\label{table:hardware}
\end{table*}

However, traditional IDM does not consider the non-negative expected distance problem when the front car moves away from the rear car at high speed. Besides, this model also ignores the perception uncertainties in human driving. Hence, we propose a \textbf{\textit{perception-uncertainty}} aware IDM for robust simulation. In this model, the acceleration speed is defined as follow:

\begin{center}
    $\ddot{x}=a\left(1-\left(\frac{v}{v_{f}}\right)^{4}-\xi\left(\Delta S^{*}(v, \Delta \mathrm{v})\right) \cdot\left(\frac{\Delta S^{*}(v, \Delta \mathrm{v})}{\Delta x}\right)^{2}\right)$
\end{center}

where $\xi\left(\Delta S^{*}(v, \Delta \mathrm{v})\right)$ is defined as below:

\begin{center}
    $\xi\left(\Delta S^{*}(v, \Delta \mathrm{v})\right)=\left\{\begin{array}{l}1 \text { if } \xi\left(\Delta S^{*}(v, \Delta \mathrm{v})\right)>0 \\ 0 \text { if } \xi\left(\Delta S^{*}(v, \Delta \mathrm{v})\right) \leq 0\end{array}\right.$
\end{center}

Besides, we introduce a perception uncertainty factor $\Delta S_p$ which is sampled from the \textit{Gaussian distribution} so that the expected follow-up distance is: 

\begin{center}
    $\Delta S^{*}(v, \Delta \mathrm{v})=S_{0}+\max \left(v \cdot \Delta t+\frac{v \cdot \Delta \mathrm{v}}{2 \sqrt{a b}}, 0\right)+\Delta S_{p}$
\end{center}

\begin{center}
    $f\left(\Delta \mathrm{S}_{\mathrm{p}}\right)=\frac{1}{\sqrt{2 \pi} \sigma} \exp \left(-\frac{\left(\Delta S_{p}-\mu\right)^{2}}{2 \sigma^{2}}\right)$
\end{center}

where $\mu$ and $\sigma$ are the mean and variance of a normal distribution.

\subsection{Hardware: Real-world Evaluation}
The SMART hardware consists of a well-designed platform ($290cm\times 190cm\times 80cm$) and a multi-robot system. It provides a physical environment close to the simulation environment, as shown in Fig.~\ref{fig:1}. Different interactive scenarios in simulation can be deployed to the platform by setting up the map and initializing the multi-robot system. To ensure that the learned policies can be executed and re-trained, we carefully developed two types of robots as follows:
\begin{itemize}
    \item Smartbot-alpha: Smartbot-alpha is composed of a mainboard handling information processing and communication, a motor board responsible for the robot's motion, and different sensors. We choose Raspberry Pi 3b+ as the mainboard in the prototype, which supports Linux, ROS, and wireless communication. To increase the robot's capability on edge computing, we use Tensor Processing Unit (TPU) produced by Google to run the deep learning models. For sensing, smartbot-alpha is equipped with a 360-degree laser scanner that can detect nearby obstacles and a 160-degree camera that can capture the environments. The laser scanner and camera can be replaced according to the requirements of different applications.
    \item Smartbot-beta: To meet the requirement of heterogeneous robots, we have developed another type of robot called smartbot-beta. Although the shapes of smartbot-alpha and smartbot-beta are very similar, there are still two differences. First, the mainboard in smartbot-alpha is Raspberry Pi 3b+, while the mainboard used in smartbot-beta is Nvidia Jetson Nano. Jetson Nano is another powerful and efficient computer that supports different AI applications. These two mainboards will cause some differences in information processing and communication. Second, the actuators of smartbot-alpha and smartbot-beta are different, leading to different motion control accuracy. Tab.~\ref{table:hardware} summarize the details of the hardware in the two robots.
\end{itemize}

\textbf{Communication and Computation.} Our platform uses wireless communication between the servers and robots. Also, the platform is well integrated with ROS and adopts the \textit{publish–subscribe} mode as the message passing method. Robots can share information with others or the server by publishing messages on a specific topic. Besides, we observe that some image processing tasks will cost lots of computation and memory. Hence, each robot will upload some image processing tasks to the server, such as extrinsic image calibration or projection.

\subsection{Supports for Training and Evaluation}
\textbf{State, Action, and Reward Adapters.} In our platform, the observation space for each robot includes different types of sensor data, such as the Lidar data, RGB image data, and robot's speed. To make the implementation easier, we provide the \textit{state adapter} to convert different data types into a NumPy array, which is a standard data type in python programming. Therefore, users can customize the state encoding in the state adapter based on their approaches. We provide both discrete action space and continuous action space for the robot's action space. Users can specify the action space in the \textit{action adapter} based on value-based and policy-based reinforcement learning \cite{zhang2019multi}. Additionally, our action adapter can convert the action data type to ROS data type that can be used to control the robots. For the reward, we provide \textit{reward shaping} function for multi-robot cooperation so that users can define the rules of reward calculation for each robot in the \textit{reward adapter}. In the autonomous driving scenarios, we provide a reward adapter that computes the forward travel distance as the positive reward for individual robots and the collision as the penalty for relative robots. 

\textbf{Dynamic Randomization.} The policy learned in simulation may achieve superior performance. However, the performance in a real-world environment can be poor because of the ubiquitous uncertainties in sensing and motion control. In the simulator, robots can observe precise sensing data while there are numerous sensor noises in the real world. Hence, we provide several dynamic randomized training techniques in the simulation include:
\begin{itemize}
    \item Randomizing the initial position of each robot;
    \item Adding Gaussian noise in the sensing data;
    \item Adding Gaussian noise in the robot's speed; and
    \item Introducing social robots controlled by the social agent model, unknown to the training robots.
\end{itemize}

\textbf{MRRL Baselines.} We classify existing popular MARL approaches into two categories: independent learning and centralized critic with decentralized policies. We implement single-robot reinforcement learning for independent learning, such as MADQN and MAPPO, and simply extend them to multi-robot cases. To investigate the non-stationary and credit assignment problems in MRRL, we provide several existing solutions based on centralized critic and decentralized policies paradigm, such as MADDPG, COMA, and VDN. To the best of our knowledge, some approaches are first evaluated in real-world environments, such as QMIX and MAAC. More details about these approaches will be introduced in Sec.~\ref{sec:case-study}.

\begin{figure}[!ht]
\centering \includegraphics[width=\linewidth]{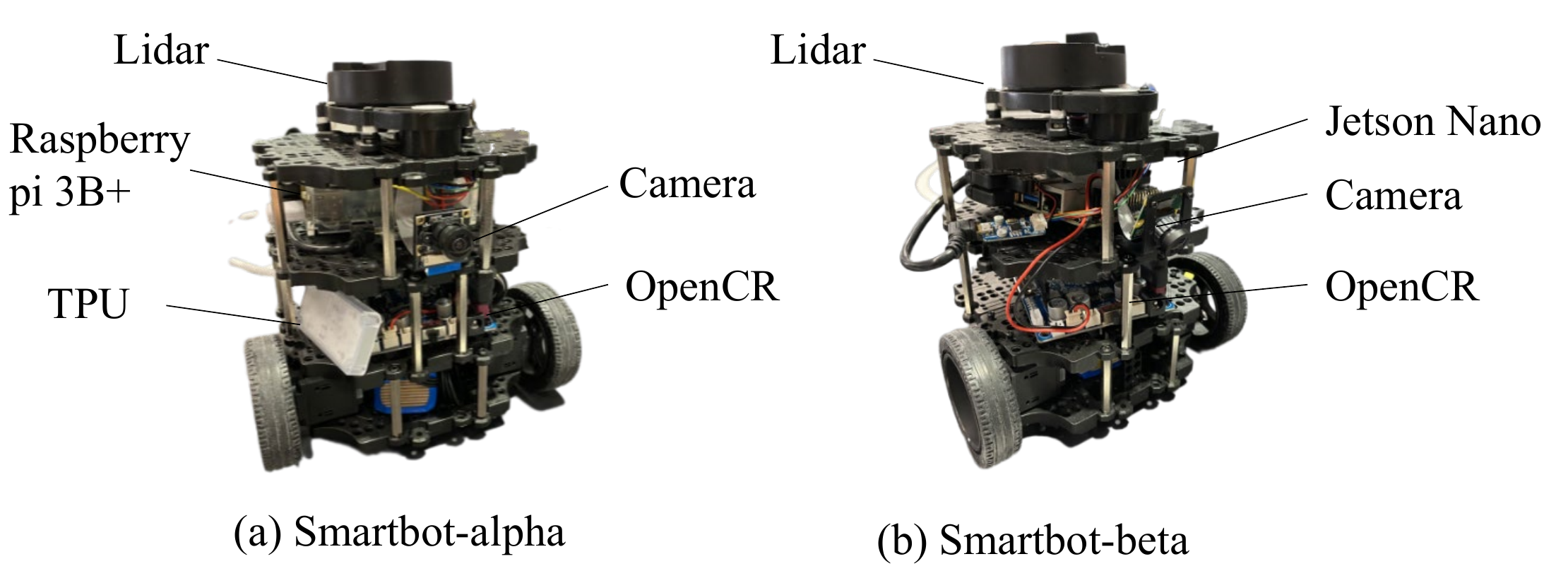}
\label{fig:robts}
\centering \caption{Left side shows our first prototype smartbot-alpha and right side is the second prototype smartbot-beta.}
\end{figure}

\begin{figure*}
\centering
\includegraphics[width=\linewidth]{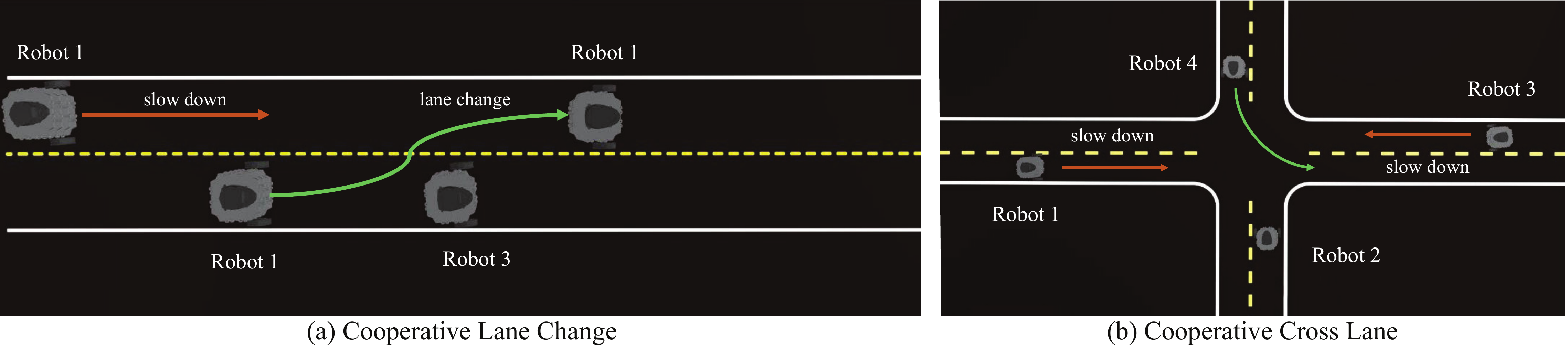}
\caption{Illustration of cooperative driving in different merge scenarios. (a) When robot $1$ performs lane change due to the close distance with front robot $3$, robot $1$ should learn to control its speed to avoid collisions. (b) Similar multi-robot interaction can also be found in cross intersection scenario}
\label{fig:lane-scenarios}
\end{figure*}

\section{Case Study: Multi-robot Cooperative Driving in Lane Merge Scenarios}\label{sec:case-study}

Autonomous driving is one of the most important applications of artificial intelligence, which has been widely studied in the past decades \cite{kiran2020deep}. Despite the remarkable achievements, some fundamental challenges remain unsolved, for example, how to learn a driving policy in a dynamic environment \cite{badue2020self}. Besides, another fundamental challenge lies in the realistic and complex cooperation among multiple autonomous vehicles. Recently, MRRL has been considered a promising approach for cooperative driving. To demonstrate how SMART may support MRRL research on cooperative driving, we select the lane merge scenario as a case study and propose three different approaches to solve it. Below, we briefly introduce MRRL approaches and experiment results of training and evaluating these approaches on SMART.

\subsection{Learning Cooperative Driving in Merge Scenarios}

\textbf{Problem Definition.} Lane change is one of the most fundamental tasks in safe autonomous driving. When the front vehicle is broken or stopped, the following vehicle needs to change to another lane to avoid accidents or traffic congestion. Fig.~\ref{fig:lane-scenarios} shows two typical cooperative lane-change scenarios . However, it is hard to define a model to control the robot movement considering the complex interaction.

MRRL provides supports to learn such cooperative policies through interaction with each other. In MRRL, multiple robots act and learn in a sharing environment. At each time, the robot receives an observation $s_i$, chooses an action $a_i$ according to its policy $\pi_i$, and receive the reward $r_i$. In cooperative MRRL, all robots share the same global reward $r_1=r_2=...=r_n$. The objective is to maximize the accumulative discount reward $R=\sum_{t=0}^{T}\gamma^t r_t$, where $\gamma \in [0,1)$ is the discount factor.

\textbf{Hierarchical Decision-making Model.} This paper proposes a hierarchical decision-making model for automated vehicle control, including high-level action selection and low-level motion control. The low-level motion control is achieved by using computer vision methods. Furthermore, we propose three different reinforcement learning-based approaches to learn a cooperative policy for high-level action selection. Instead of learning a low-level cooperation strategy, learning high-level policy can reduce the complexity of learning cooperation. The followings are the specific definition of state, action, and reward in the cooperative lane change task.

\textbf{State Space}. For each robot, the state is the information used to make decisions. Many existing cooperative lane-change models assume that each robot can obtain the dynamic position of other robots. In practice, it is non-trivial to know the position of other robots. In this paper, we consider a more challenging situation where each robot can only obtain the information from its observation $s_{o}$ and the attributes of the robot itself $s_{a}$. The state of each robot is defined as follows:

\begin{center}
    $s=[s_{o},s_{a}]$
\end{center}
where $s_{o}$ is the real-time sensing data from lidar and $s_{a}$ can be the speed and lane flag.

\begin{figure*}[!ht]
\centering
\includegraphics[width=\linewidth]{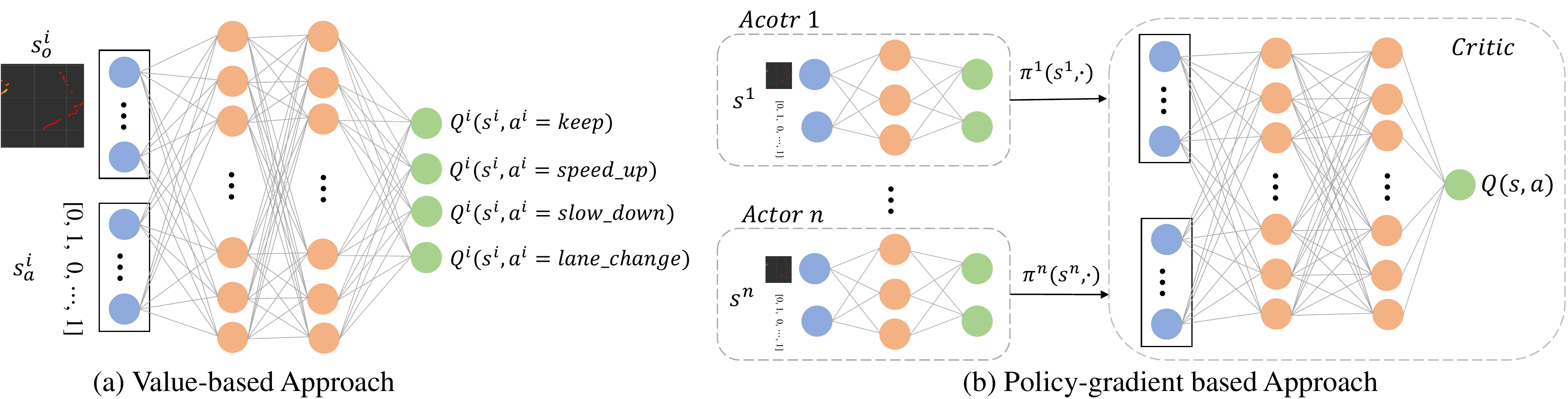}
\caption{Illustration of the different deep neural network architecture of value-based and policy-gradient based reinforcement learning approaches}
\label{fig:models}
\end{figure*}

\textbf{Action Space}. SMART provides discrete and continuous action spaces for training. Here, we consider a discrete action space for simplicity and the action space of each robot is:
\begin{center}
    $a={\left \{slow\_down, keep\_lane, speed\_up, left\_change \right \}}$
\end{center}
The robot can select to slow down, keep current speed, speed up, or lane change at each time step. For example, $a=[0 0 1 0]$ denotes the robot select to speed up. Note that the selected action will be represented as a one-hot vector and will be transferred to Twist message by our action adapter to control the vehicle through ROS.

\textbf{Reward Design}. Typically, safety and efficiency are two main concerns in cooperative merge scenarios. In terms of safety, each vehicle should avoid collision with the front vehicle when driving in the current lane and pay attention to the vehicles in the merged lane. Once the collision happens, we will assign a negative reward to the robots. In particular, a penalty of $r_{col}$ will be given to the robot, and the current training episode will end and restart. To avoid traffic congestion, we would like to encourage the vehicle to run as fast as possible, and a positive reward $r_{travel}$ will be given to vehicles. Hence, the total reward of each vehicle is designed as follows:

\begin{center}
    $r_{total}=\alpha r_{col}+(1-\alpha)r_{travel}$
\end{center}

where $r_{col}$ is the penalty of collision and $r_{travel}$ is the reward for moving forward. Here, $\alpha$ is a hyperparameter to control the weight of collision avoidance and move forward.

\begin{table*}[!ht]
\centering
\caption{A comparison of difference MARL approaches evaluated in the case study}
\begin{tabular}{|c|c|c|c|c|}

\hline 
 & Base Algorithms & \makecell[c]{How to Measure Rach Robot's \\ Contribution to Team Performance} & Number of Critics & Exploration Strategy \\
\hline 
IDQN & DQN & Shared Team Rewad & N & $\epsilon$-greedy \\
\hline
MADDPG & DDPG & Shared Team Rewad & $1$ & OU Noise \\
\hline
COMA & Actor-critic & Counterfactual Inference & $1$ & OU Noise \\
\hline
VDN & DQN & Value Decomposition & $N$ & $\epsilon$-greedy \\
\hline
QMIX & DQN & Value Decomposition & $N$ & $\epsilon$-greedy \\
\hline
MAAC & SAC & Individual Rewad Design & $N$ & Maximum Entropy\\
\hline
\end{tabular}

\end{table*}

\subsection{Multi-robot Reinforcement Learning Approaches}

\textbf{Independent Deep Q-learning (IDQL).} Deep Q-Networks (DQN) are the most popular deep reinforcement learning approach which aims to learn the action-value $Q^{\pi}(s, a)=\mathbb{E}_{s^{\prime}}\left[r(s, a)+\gamma \mathbb{E}_{a^{\prime} \sim \pi}\left[Q^{\pi}\left(s^{\prime},a^{\prime}\right)\right]\right]$. Here, we directly extend it to multi-robot cooperation and the loss function for training Q-network is:
\begin{center}
    $\mathcal{L}_i(\theta)=\mathbb{E}_{s_i, a_i, r, s_{i}^{\prime}}\left[\left(Q_i\left(s_{i},{a}_{i};\theta\right)-y_i\right)^{2}\right], \quad$
\end{center}
where $y_i=r+\gamma \max _{{a_i} \in \mathcal{A_i}}\left(Q_i\left({s}_{i}^{'}, {a_i};\theta_i^{-}\right)\right)$ and $\theta_i^{-}$ is the \textit{target network} that are periodically copied from $\theta_i$ and kept constant for a number of iterations.

\textbf{Value Decomposition in Multi-robot Q-learning.} Naive independent Q-learning ignores the mutual influence between robots, which may cause non-stationary issues during learning. One way to address the issue is to learn a jointed action-value function that considers all robots' states and actions, called a jointed action learner (JAL). Similarly, the loss function of jointed action learner is:

\begin{center}
    $\mathcal{L}(\theta)=\mathbb{E}_{\bm{s}, \bm{a}, r, \bm{s^{\prime}}}\left[\left(Q\left(\bm{s},\bm{a};\theta\right)-y\right)^{2}\right], \quad$
\end{center}

where $\bm{s}=[s_1,s_2,...,s_n]$ and $\bm{a}=[a_1,a_2,...,a_n]$ are the joint state and joint action of all robots, respectively. On the one hand, JAL suffers from the curse of dimensionality during training and execution because the action space will expand dramatically when the number of robots increases. On the other hand, JAL cannot distinguish each robot's contribution to the team's success. Several value decomposition methods are proposed to address the issues. In VDN, the global Q-value is approximated by the sum of individual robot's Q-value:

\begin{center}
    $Q_(s,a)=\sum_{i}^{N}Q_{i}(h_i,a_i)$
\end{center}

where $h_i$ is the hidden state of each robot's and $Q_i$ denotes individual robot's Q-value. Similarly, QMIX proposed a monotonic and non-linear method to factorize the global Q-value by introducing a mixed network and monotonic constraint:

\begin{center}
    $$\frac{\partial Q(s,a)}{\partial Q_{i}(s_i,a_i)} \geq 0$$
\end{center}

The proposed monotonic constraint ensures consistency between the centralized and decentralized policies.

\begin{figure*}[!ht]
\centering
\includegraphics[width=\linewidth]{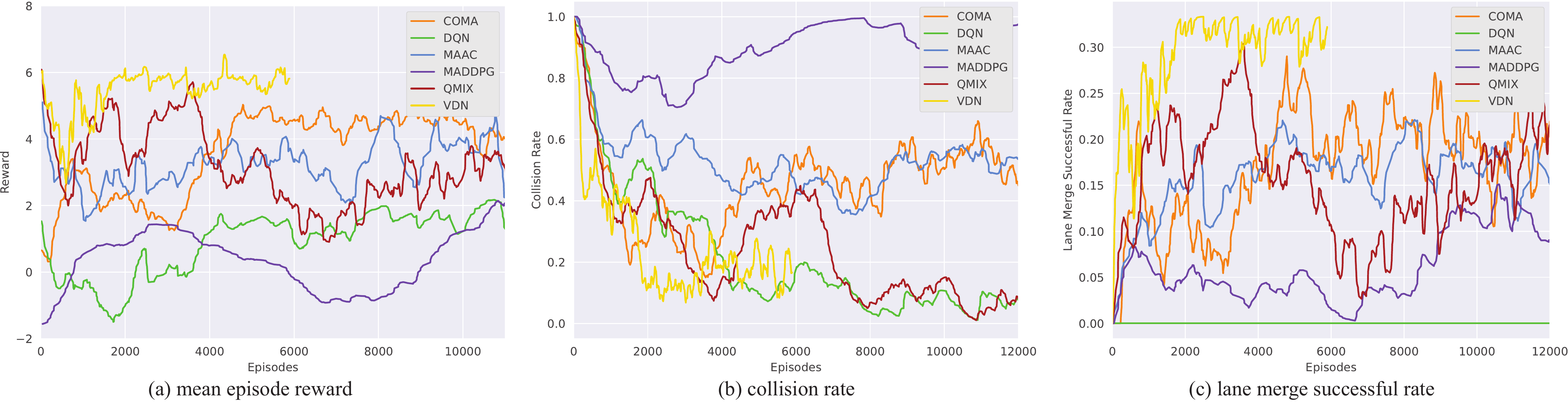}
\caption{Comparison of the learning curve of different approaches in the cooperative lane change scenarios}
\label{fig:lane-change-learning-curve}
\end{figure*}

\textbf{Policy Gradient and Actor-Critic.} Both independent DQN and joint action learner are value-based methods where the policy is derived by selecting the action with the highest $Q$ value. An alternative model-free approach is the policy-based method which directly learns a policy $\pi$ parameterized by $\theta^{\pi}$. The objective of these approaches are to adjust the parameters $\theta^{\pi}$ in order to maximize the function $J(\theta)=\mathbb{E}_{s \sim p^{\pi}, a \sim \pi_{\theta}}[R]$, where $R$ denotes the accumulative expected rewards. The term $R$ is in the policy gradient estimator leads to high variance, as these returns can vary drastically between episodes. Actor-critic methods aim to address this issue by using a function $Q(s, a)$ to approximate the expected returns, and replacing the original return term in the policy gradient estimator. The gradient of each robot's policy can be written as:

\begin{center}
    $\nabla_{\theta_{i}} J\left(\theta_{i}\right)=\mathbb{E}_{s \sim p^{\mu}, a_{i} \sim \pi_{i}}\left[\nabla_{\theta_{i}} \log \boldsymbol{\pi}_{i}\left(a_{i} \mid s_{i}\right) Q_{i}^{\boldsymbol{\pi}}\left(\mathbf{s}, a_{1}, \ldots, a_{N}\right)\right]$
\end{center}

Then, the joint action-value function can be trained independently with TD-learning and parameter sharing.

\textbf{Counterfactual Multi-robot Policy Gradient (COMA).} Foerster et al. extended the actor-critic method to a multi-agent setting and proposed a \textit{counterfactual baseline} to infer each robot's contribution \cite{foerster2017counterfactual}. The method trains a centralized critic to compute an agent-specific \textit{advantage function} $A(s,a)$ by marginalizing out a single agent's action while keep other agents' action fixed:

\begin{center}
    $A(s, a)=Q(s, a)-\sum_{a_{i}^{\prime}} \pi_{i}\left(a_{i}^{\prime} \mid o_{i}\right) Q\left(s,\left(a_{-i}, a_{i}^{\prime}\right)\right)$
\end{center}

where $a$ is the action of all robots and $a_{-i}$ denotes the other robots' actions except robot $i$. Then, the gradient used to update each robot's policy network can be computed as:

\begin{center}
     $\nabla_{\theta_{i}} J\left(\theta_{i}\right)=\mathbb{E}_{s \sim p^{\mu}, a_{i} \sim \pi_{i}}\left[\nabla_{\theta_{i}} \log \boldsymbol{\pi}_{i}\left(a_{i} \mid s_{i}\right) A_{i}\left(\mathbf{s},a_{i}\right)\right]$
\end{center}
Also, we argue that this counterfactual baseline has better interpretability to measure individual robots' contributions to team performance.

\begin{table}[!ht]
\centering
\caption{Hyperparameters for Training}
\begin{tabular}{|c|c|}

\hline 
\textbf{Hyperparameter} & \textbf{Value}\\
\hline 
Training Episode & $30,000$\\
\hline
Episode Length & $30$\\
\hline
Buffer Capacity & $100,000$\\
\hline
Batch Size & $1024$\\
\hline
Learning Rate & $0.01$\\
\hline
Discount Factor $\gamma$ & $0.95$\\
\hline
Hidden\_dim & $32$\\
\hline
Target Network Update Rate & $0.01$\\
\hline

\end{tabular}
\label{tab:hyperparameter}
\end{table}

\textbf{Actor-attention-Critic Method (MAAC).} The attention mechanism is emerging as a successful approach to select contextual information, which has been widely applied in computer vision and natural language processing. Iqbal et al. applied an attention mechanism to train the critic by dynamically selecting the most relevant information from other robots at each time \cite{islam2020hamlet}. When training the critic, the information from other robots can be computed as follows:

\begin{center}
    $x_{i}=\sum_{j \neq i} \alpha_{j} v_{j}=\sum_{j \neq i} \alpha_{j} h\left(V g_{j}\left(o_{j}, a_{j}\right)\right)$
\end{center}

where $v_j$ is each robot's embedding, $g_j$ is multi-layer perceptron (MLP), and $o_j$ denotes the observation of robot $j$. Besides, $V$ is a shared matrix and $h$ is an element-wise nonlinearity. The attention weight is calculated as follow:

\begin{center}
    $\alpha_{j} \propto \exp \left(e_{j}^{\mathrm{T}} W_{k}^{\mathrm{T}} W_{q} e_{i}\right)$
\end{center}

where $W_q$ transfer the $e_i$ into ``query'' and $W_k$ transfer the $e_j$ into ``key''. Then, the $Q$ function of each robot $i$ is represented as:

\begin{center}
    $Q_{i}(o, a)=f_{i}\left(g_{i}\left(o_{i}, a_{i}\right), x_{i}\right)$
\end{center}

After that, each robot's policy is updated by ascent with the following gradient:

\begin{center}
    $\nabla_{\theta_{i}} J\left(\pi_{\theta}\right)=$
$\mathbb{E}_{o \sim D, a \sim \pi}\left[\nabla_{\theta_{i}} \log \left(\pi_{\theta_{i}}\left(a_{i} \mid o_{i}\right)\right)\left(-\alpha \log \left(\pi_{\theta_{i}}\left(a_{i} \mid o_{i}\right)\right)+\right.\right.$
$\left.\left.Q_{i}(o, a)-b\left(o, a_{\backslash i}\right)\right)\right]$
\end{center}

where $-\alpha \log\left(\pi_{\theta_{i}}\left(a_{i} \mid o_{i}\right)\right)$ is used to encourage exploration and avoid converging to non-optimal deterministic policies and $\left.b\left(o, a_{\backslash i}\right)\right)=\mathbb{E}_{a_{i} \sim \pi_{i}\left(o_{i}\right)}\left[Q_{i}\left(o,\left(a_{i}, a_{\backslash i}\right)\right)\right]$ inspired by the advantage function.

\textbf{Evaluation Metrics} In this paper, we consider four evaluation metrics to measure the performance of different approaches, which are:
\begin{itemize}
    \item \textbf{Mean episode reward}: we sample the replay buffer from the different episodes and calculate the mean reward of each time step;
    \item \textbf{Collision rate}: the collision radio among robots merge in each episode;
    \item \textbf{Lane merge successful rate}: the proportion of robots who successfully merge in each episode; and
    \item \textbf{Mean speed}: the mean of each robot's speed at each time step.
\end{itemize}

We apply the above methods to solve the cooperative lane merge problem and the model architecture of value-based methods and policy-gradient methods as illustrated in Fig.~\ref{fig:models}. To the best of our knowledge, some of them are first applied to train and evaluation in multi-robot cooperative driving scenarios, such as QMIX and MAAC.

\subsection{Training in Simulator}

\begin{figure*}[!ht]
\centering
\includegraphics[width=\linewidth]{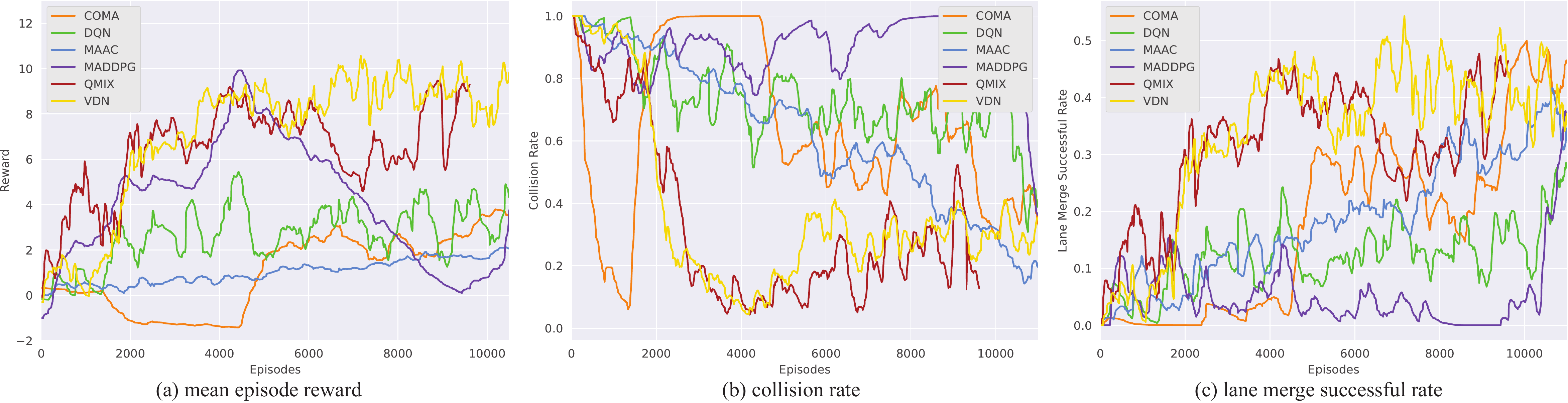}
\caption{Comparison of the learning curve of different approaches in the cooperative cross lane scenario}
\label{fig:cross-lane-learning-curve}
\end{figure*}

We conduct two experiments in the SMART simulation environment and initialize the environment with a cooperative lane change and cooperative cross lane scenarios. In the first scenario, robot $3$ is set to static, and robot $2$ needs to control itself to perform lane change which needs coordination among robots $1$ and $3$. Four robots must pass the intersection in the second scenario without colliding with others. At the beginning of each episode, we randomly initialize several positions of the robots and set the same initial speed to them in these two scenarios. We set the episode length to be $18$ and $24$ in the cooperative lane change and cooperative cross lane scenarios, respectively. The hyperparameters of our network architecture and learning algorithms are presented in Tab.~\ref{tab:hyperparameter}. When the vehicles get collided, or the max episode length is reached, the program will call the \textit{reset} APIs provided by the SMART to reset the environment and continue the training.

As shown in Fig.~\ref{fig:lane-change-learning-curve}(a), several approaches, such as COMA, VDN, and QMIX, show a slight decrease at the beginning due to the exploration in the action space and then gradually grow up. Besides, VDN shows a faster converge rate than the other approaches. Fig.~\ref{fig:lane-change-learning-curve}(b) and Fig.~\ref{fig:lane-change-learning-curve}(c) indicate the changing of collision rate and lane merge successful rate, respectively. All of MAAC, VDN, and DQN reach a low collision rate of almost $0.02$ after $8000$ episodes. However, DQN achieves a zero lane merge successful rate. We check the performance of DQN in the Gazebo simulation and found that robot $2$ learns a strategy to keep the minimum speed instead of changing the lane. This strategy can also avoid the collision with robot $3$ within the $18$ episode length.
Moreover, we observe that the value-based methods have a faster coverage rate than the policy-gradient ones. The different convergence rates may come from the different strategies of action exploration. Most value-based methods use $\epsilon$-greedy exploration strategy while the policy-gradient methods leverage OU noise and maximum entropy.

The interaction among robots in the cross-lane scenario is more complex than the first scenario in terms of the number of robots and the need for lane merge. Fig.~\ref{fig:cross-lane-learning-curve}(a) shows the trend of mean episode reward during the training. As we can see, VDN and QMIX show a faster coverage speed than the others. Also, the learning of MAAC is prolonged, and the learning curve of MADDPG is non-stationary. In Fig.~\ref{fig:cross-lane-learning-curve}(b), MAAC significantly decreases during the training and reaches a low collision rate closer to VDN and QMIX. Fig.~\ref{fig:cross-lane-learning-curve}(c) indicates that VDN, QMIX, and MAAC can achieve a higher lane merge success rate, which is near $0.5$ when the episode length is set to $24$.

\subsection{Evaluation with Social Agent}

\begin{figure}[!ht]
\centering
\includegraphics[width=\linewidth]{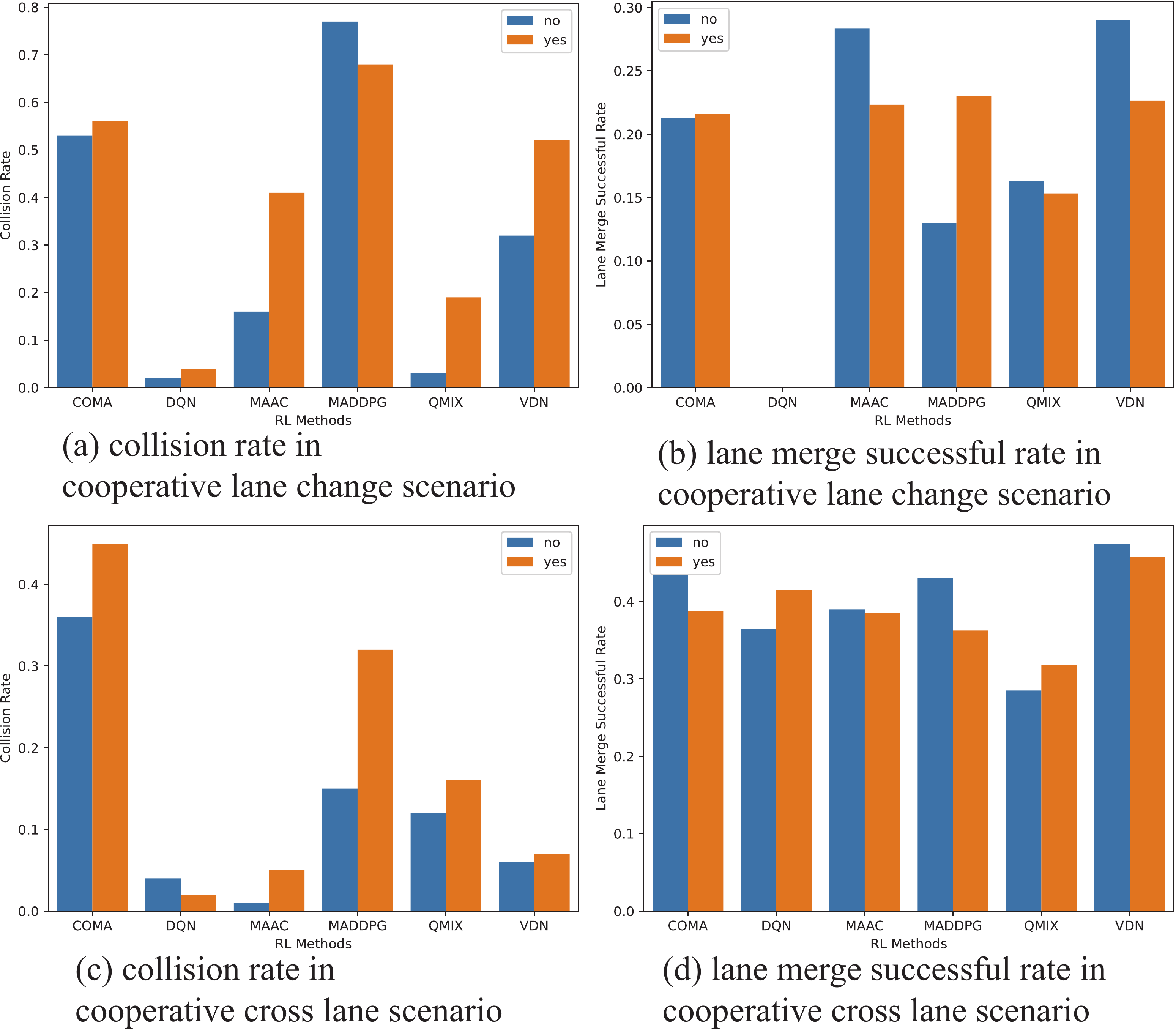}
\caption{Performance of different approaches when interacting with the social agent.}\label{fig:bar-performance}
\end{figure}

We randomly replace one robot with the social agent model that we proposed to evaluate a multi-robot system's robustness and fault tolerance. Besides, we extend the training episodes of MADDPG and MAAC to obtain the best performance of their models. As we can see from Fig.~\ref{fig:bar-performance}, the collision rate of most evaluated methods increases after introducing the social agent. It means that the existing MRRL approaches are not robust to environmental changes. Such an issue is rarely discussed in the current reinforcement learning research. Besides, we observe that MAAC performs very well in the complex cross-lane scenario as it maintains a low collision rate and lane merge successful rate. The robustness problem of MRRL will be discussed later.

\subsection{Evaluation in the Real-world platform}

\begin{figure*}[!ht]
\centering \includegraphics[width=\linewidth]{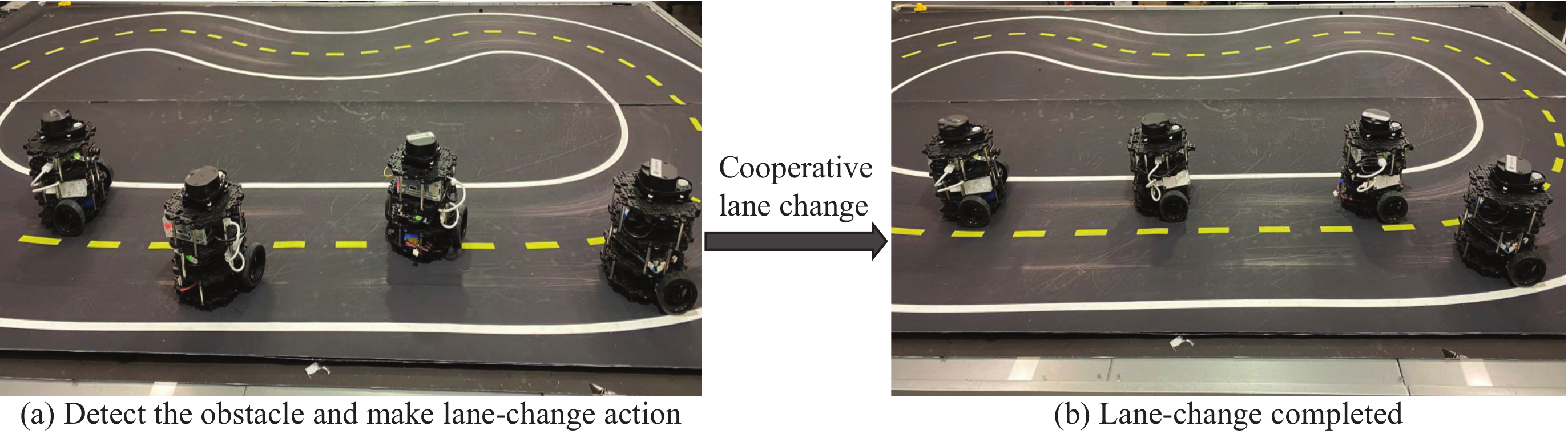}
\caption{Real-world evaluation}
\label{fig:real-world-eva}
\end{figure*}

\begin{table}[!ht]
\centering
\caption{Performance evaluation in real-world platform}
\begin{tabular}{|c|c|c|c|}
\hline
\diagbox{Method}{Metrics} & Collision Rate & \makecell[c]{Successful \\ Rate} & \makecell[c]{Mean \\ Speed} \\
\hline 
COMA & $0.35$ & $0.65$ & $0.06344$ \\
\hline 
DQN & $1$ & $0$ & $0.05395$ \\
\hline
MAAC & $0.15$ & $0.85$ & $0.05571$ \\
\hline
MADDPG & $0.95$ & $0.5$ & $0.07029$ \\
\hline
QMIX & $0.5$ & $0.5$ & $0.05858$ \\
\hline
VDN & $0.2$ & $0.8$ & $0.05498$ \\
\hline
\end{tabular}
\label{tab:real-performance}
\end{table}

\textbf{Configuration}. To investigate the gap between simulation and reality, we deploy the learned policies in simulation to the real-world multi-robot system and run $24$ episodes for each MRRL method with several random initial positions. Besides, we run a master node on the server to monitor the state of each robot and calculate the collision rate, lane merge successful rate, and mean speed. Furthermore, each robot can know the state of the team and the reward can be computed using the reward function proposed in section \uppercase\expandafter{\romannumeral3} A.

As shown in Tab.~\ref{tab:real-performance}, MAAC and VDN reach lower collision rates than others, which are $0.15$ and $0.2$, respectively. Although DQN achieves a low collision rate in the simulation, it performs worst in real-world testing. Robot $1$ fails to slow down to coordinate with robot $2$. By contrast, MAAC achieves a meager collision rate in the real world. The reason is that MAAC adopts entropy maximization in training, which can obtain diverse and as stochastic as possible strategies.

\section{The Lessons We Learned}\label{sec:lesson}
This section summarizes the lessons we learn from the case study. The target is to present the reader with a high-level summary of the capabilities of current MARL methods in the multi-robot system domain, discuss which issues make deployment of MARL methods difficult, and provide perspectives on how the difficulties can be mitigated.

\subsection{Training Mechanism of MRRL}
One fundamental challenge in real-world MRRL is the training mechanism. The most reliable and stable training mechanism is pre-training in the simulation. There are several reasons why simulation is essential to robotic learning. First, real-world training may need to \textit{reset} the environment, such as reset the initial positions of robots. This reset process will require an additional location system and be time-consuming. Second, the robots need to be recharged after running for a while. By contrast, it is easy to reset and train in the simulation environment. Moreover, the unsafe action exploration due to ``trial-and-error'' may cost undesirable damage to the robots.

\subsection{Sampling Efficiency}

Sampling efficiency is another challenging problem in real-world MRRL. Unlike supervised learning, reinforcement learning collects the experience (i.e., data) from the interaction with the environment. The high dimensional state and action spaces result in millions of environmental interactions in robotic learning. As shown in Fig.~\ref{fig:cross-lane-learning-curve}(a), the learning process of MAAC is very slow, while the learning curve of MADDPG shows significant vibration during training. Without any improvements on the algorithms, the number of training steps will increase when the model size increases or the task becomes complex. One feasible solution might be imitation learning which pre-train the reinforcement learning policy with demonstration data \cite{chi2020collaborative}.

\subsection{Robustness of MRRL}

A centralized critic with a decentralized actor (CCDA) has been a popular training paradigm in MARL and MRRL. In this method, a centralized critic network is trained to estimate the global Q-value while each robot maintains a decentralized actor network. We argue that the centralized value function enforces each robot to form a consensus strategy to perform the tasks, reducing the non-stationary training problem. However, the strongly cooperative strategy may not be robust to other robots' changing or failing. As shown in Fig.~\ref{fig:bar-performance}, the collision rate increases when we randomly replace one robot with a social agent model. Besides, the robustness and fault-tolerant in MRRL are not often the focus of mainstream reinforcement learning research.

\subsection{The Gap between Simulation and Reality in MRRL}

Although simulation provides a cheaper, faster, and safer way to train the robots than real-world experiments, it brings several challenges. First, the state space and the transition probability are not the same as reality, especially in the robotic area. It is non-trivial to simulate the physical mechanism of robots and the real-world model. Hence, we provide different adapters to support reshaping the state space when training the algorithms. Second, traditional MDP formulation assumes synchronous execution, where the state remains unchanged until the action is applied and all the robots take actions synchronously. We argue that this assumption is not suitable for multi-robot systems. Finally, the real-world environment also involves many uncertainties, such as sensor noise and communication latency.

\section{Related Work}\label{sec:related-work}
\begin{table*}[!ht]
\centering
\caption{Summary of existing evaluation platforms in reinforcement learning research}
\begin{tabular}{|c| c |c |c |c| c|}
\hline
\textbf{Platform} & \textbf{Scenario} & \textbf{Environment} & \textbf{Action Space} & \textbf{Language} & \textbf{Limitations}  \\
\hline
ALE & Game & 2D & Discrete &  C++ & Hard to compile and install \\
\hline
OpenAI Gym &  Game & 2D & Discrete & Python & Lack of physics engine\\
\hline
Particle Environment &  Game & 2D & Discrete/Continuous & Python & Limited physics simulation\\
\hline

TORCS &  Game & 3D & Discrete/Continuous & Python & Focus on racing scenarios\\
\hline
SC2LE(SMAC) &  Game & 3D & Discrete/Continuous & Python & Lack of physics engine\\
\hline
MuJoCo soccer &  Game & 3D & Discrete/Continuous & Python & Lack of diversity complex interaction\\
\hline
MARL-Ped &  Traffic Flow & 3D & Continuous & C++ & Hard to compile and install\\
\hline
PTV VISSIM &  Traffic Flow & 2D/3D & Discrete/Continuous & C++/Python & Not completely open source\\
\hline
SUMO &  Traffic Flow & 2D & Discrete/Continuous & C++/Python & Low-fidelity simulation\\
\hline
Traffic Flow \cite{zhang2019cityflow} &  Traffic Flow & 2D & Discrete/Continuous & C++/Python & Low-fidelity simulation\\
\hline
Gazebo &  Autonomous Driving & 3D & Discrete/Continuous & Python & Need to provide the environment model\\
\hline
\end{tabular}

\end{table*}
MRRL is an interdisciplinary research topic with a long history. MRRL research varies from game theory, machine learning, and stochastic control to optimization. Classic MRRL has been widely studied in zero-sum games and extensive-form games where games can be represented as a matrix. The rows of the matrix correspond to the action set of the first player, while the columns denote the action set of the second player. The values of the matrix are the payoffs of a given joint-action pair. In this scenario, evaluating different MRRL approaches is to find the nash equilibrium from the matrix.

Arcade Learning Environment (ALE) was introduced after deep reinforcement learning showing its potential to solve various optimal control problems \cite{bellemare2013arcade}. It provides an interface to hundreds of Atari $2600$ game environments, including both single-robot and multi-robot environments. However, the ALE was initially challenging to compile and install because it involves $C$ APIs and an unofficial fork with a python wrapper. There were different benchmark environments for independent projects in various languages and unique APIs. Doing simple researches in reinforcement learning requires an organization with software engineering divisions. Therefore, OpenAI Gym was created to promote research in reinforcement learning by making comprehensive benchmarking much more accessible \cite{brockman2016openai}. It integrated multiple environments into one simple Python APIs that the researcher could easily understand. At the same time, many similar environments and reinforcement learning libraries were proposed, including competitive and cooperation scenarios. In these scenarios, robots interacted with each other in a two-dimensional world with continuous action space and discrete-time. The state-space in these environments only involves different robots, so it can not represent the complex environment. 

SC2LE (StarCraft II Learning Environment) offered a more complex environment based on the game StarCraft II \cite{vinyals2017starcraft}. It proposes a challenge for reinforcement learning and MRRL where robots play with others in a partially observable map \cite{wang2020qplex}. In this scenario, heterogeneous robots can only observe other robots if they are alive and within the sight \cite{wang2020roma}\cite{wang2019learning}. The feature of the robots' state contains attributes of the robot itself and its neighbors such as relative position, health, shield, and types. Although SC2LE provides a complex interaction environment, it does not offer the physics engine vital for real-world multi-robot cooperation.

Several simulators are developed to simulate the physics of the real-world environment and the dynamics of robotic control to meet this requirement in the robotics area. MuJoCo soccer provides a challenging competitive multi-robot soccer environment with continuous simulated physics \cite{liu2019emergent}. In MuJoCo soccer, different soccers (robots) play in a  wider universe of possible environments with consistent simulated physics. Each robot needs to control different body parts and cooperate with others. It raises the challenge of learning robot locomotion and collaboration. Gazebo is another popular lightweight physics engine including convenient programmatic and graphical interfaces for implementation \cite{chen2019mapless}\cite{gym-pybullet-drones2020}. Specifically, Gazebo provides python APIs to interact with the robot operating system (ROS), which is widely used in real-world multi-robot systems \cite{quigley2009ros}. According to their application scenarios, the researcher can build their world and robot models in the simulator. Similar 3D physics engines can be found in CARLA and Airsim. To be sure, there are some real-world multi-agent system investigating the interaction among multiple agents \cite{pickem2017robotarium}\cite{wang2019pattern}\cite{li2021group}\cite{yang2020multiplexity}. Recently, Xiao et al. developed a reinforcement learning-based physical-Layer authentication for controller area networks such as telematics ECUs \cite{xiao2021reinforcement}. However, these platforms are not MRRL-friendly to the training and evaluation as they neither provide the simulation environments nor the multi-agent reinforcement learning baselines. Furthermore, the application of MARL in the multi-robot system and the gap between simulation and reality are still unclear and demand investigations.

\section{Conclusion}\label{sec:conclusion}

Multi-robot reinforcement learning is an attractive research topic as it provides a promising approach to learn from the interaction with the environment. In this paper, we discuss the training and evaluation environment in MRRL and introduce the SMART platform. SMART aims to build a scalable learning platform that can be used to evaluate different MRRL approaches. To show the support of SMART, we conduct a case study on two cooperative merge scenarios and propose three different approaches. These approaches are trained in the SMART simulation environment with reinforcement learning-friendly APIs and further evaluated in the real-world platform. Based on the case study, we summarize the lessons we learned and discuss several challenges in multi-robot reinforcement learning research. Moreover, We plan to open-source the realistic simulation environments and design of our platform to encourage further research on multi-robot reinforcement learning.

\bibliographystyle{IEEEtran}
\bibliography{main}

\end{document}